\documentclass[useAMS,usenatbib,times]{mn2e}
\usepackage{amsmath}                                                            
\usepackage{amsfonts}
\usepackage{graphicx}
\usepackage{float}
\usepackage{morefloats}
\usepackage{listings}

\title[A Particle-By-Particle M2M algorithm]
{Disc galaxy modelling with a particle-by-particle M2M method}
\author[J. A. S. Hunt and  D. Kawata]
 {Jason A. S. Hunt,$^{1}$\thanks{E-mail: jash2@mssl.ucl.ac.uk}
Daisuke Kawata,$^{1}$\thanks{E-mail: dka@mssl.ucl.ac.uk}
\\
$^{1}$ Mullard Space Science Laboratory, University College London,
Holmbury St. Mary, Dorking, Surrey, RH5 6NT, UK
\\
}
\date{Accepted 2013 January 2.  Received 2012 December 11; in original form 2012 July 27 }

\pagerange{\pageref{firstpage}--\pageref{lastpage}}
\pubyear{2012}

\begin{document}

\maketitle

\label{firstpage}

\begin{abstract}
We have developed the initial version of a new particle-by-particle adaptation of the made-to-measure (M2M) method, aiming to model the Galactic disc from upcoming Galactic stellar survey data. In our new particle-by-particle M2M, the observables of the target system are
compared with those of the model galaxy at the position of the target stars (i.e. particles). The weights of the model particles are changed to reproduce the observables of the target system, and the gravitational potential is automatically adjusted by the changing weights of the particles. This paper demonstrates, as the initial work, that the particle-by-particle M2M can recreate a target disc system created by an $N$-body simulation in a known dark matter potential, with no error in the observables. The radial profiles of the surface density, velocity dispersion in the radial and perpendicular directions, and the rotational velocity of the target disc are all
well reproduced from the initial disc model, whose scale length is different from that of the target disc. We also demonstrate that our M2M can be applied to an incomplete data set and recreate the target disc reasonably well when the observables are restricted to a part of
the disc. We discuss our calibration of the model parameters and the importance of regularization.
\end{abstract}

\begin{keywords}
methods: N-body simulations --- galaxies: structure
--- galaxies: kinematics and dynamics --- The Galaxy: structure 
\end{keywords}

\section{Introduction}
\label{intro-sec}

There is still a gulf between our theoretical galaxy models and the observational data, that must be bridged before we can have a fully dynamical model of the Milky Way which is consistent with its observed properties. The major developments have been localised to certain regions of the Milky Way and the structure of many other regions of the Milky Way remains largely uncertain. 

For the last two decades Galactic astronomy has been relying on Hipparcos data \citep[e.g.][]{P97}. However, new space-based astrometry missions are going ahead in the near future, and ground-based surveys, e.g. PanStaars, \citep[e.g.][]{Ketal10}, VISTA \citep[e.g.][]{Metal09}, LSST \citep[e.g.][]{Ietal08}, SEGUE \citep[e.g.][]{Yea09}, APOGEE \citep[e.g.][]{APetal08} and RAVE \citep[e.g.][]{SEA06}, will add significant value to these missions, which will expand our knowledge of the Milky Way. The next space based surveys are Nano-JASMINE and ESA's cornerstone mission, Gaia. Nano-JASMINE is a demonstration mission, but it will likely improve on Hipparcos' proper motions. Gaia is expected to launch in 2013 and will operate for five years, with a possible one or two year extension. Gaia will provide an unprecedentedly large amount of information with which to build a more accurate model of the Milky Way.

Constructing accurate models of the Milky Way is important for allowing us to understand and compensate for observational bias, which are present in all existing Galactic surveys due to dust, gas and our location within the disc. They also allow us to tie together data from different surveys, assembling them into a single model.
There are three different types of galaxy model. Mass models only describe the density distribution and the galactic potential \citep[e.g.][]{KZS02}. Kinematic models specify the density and velocity distributions, but lack the constraint that the model must be in a steady state in the galactic potential \citep[e.g.][]{Retal04}. A model which also satisfies this criterion is known as a dynamical model \citep[e.g.][]{WPD08}. There are arguably five different types of dynamical galaxy model, although sometimes where the line of distinction is drawn can be ambiguous. 

Moment based methods find solutions of the Jeans equation that best fit the observed moments and minimise $\chi^{2}$ \citep[e.g.][]{Y80,BDI90,MB94,M95,C08,Cetal09}. The main drawback of this method is that there is no guarantee that there will be a positive distribution function with the required velocity moments. It is also usually restricted to spherically symmetric models as the symmetry allows simplified assumptions to be made. Distribution function based methods fit the distribution function $f(\textbf{r},\textbf{v})$ to the data directly. The methods have been applied to spherical or integrable systems \citep[e.g.][]{D84,B87,G91,HQ93,MT94,K95,M96,dBLD00}. Perturbation theory can be used to extend the method to near integrable potentials \citep[e.g.][]{MG99,B10}. Schwarzschild's method works by computing a large number of orbits evolved over many orbital periods in a fixed potential. Information is collected in an orbit library, and they are weighted to produce the best fit to the target model \citep[e.g.][]{S79,S93,CdZvdMR99,Getal03,Ketal05,Cetal06,Tetal09}. This method has the advantage of not requiring the distribution function or the other integrals of motion, and rarely, the distribution function may even be recovered \citep{HEDB00}. This method is not restricted by symmetry, but due to complexity it is usually only used for axisymmetric models. Recently \cite{vdBetal08} have developed a triaxial Schwarzschild method and applied it to NGC 4365. Torus methods are very similar to orbit based methods, and are often labelled within the same category. The key difference between torus modelling and orbit based modelling is that while in orbit based modelling, the orbits are time series of phase-space points, in torus modelling, these are replaced by orbital tori \citep[e.g.][]{McB12,B12-2,B12-1}. For a more detailed explanation, and a list of advantages of torus methods over orbit methods, see \cite{McB11}. Finally $N$-body models, which are the simplest to construct, are based on gravitational attraction between `$N$' bodies and can be collisional, or collisionless. The key assumption of stellar dynamics in the Galaxy is that these stellar systems are collisionless, hence collisionless $N$-body models are good approximations for galactic dynamics \citep[e.g.][]{DR11}. 

We will demonstrate an $N$-body method that allows us to recover a model of the desired galaxy, with some flexibility on the initial conditions. Our method is based upon the original made-to-measure (M2M) method by \cite{ST96} which is capable of constructing $N$-body equilibrium systems by maximising a linear combination of the entropy, and minimising $\chi^{2}$, the the mean-square deviation between the observables and the model. The M2M algorithm has been improved upon by \cite{DeL07}, \cite{Deh09}, \cite{LM10} and \cite{MG12} and has been used for a variety of tasks, as detailed below. \cite{BDG04} apply the M2M algorithm from \cite{ST96} to the Milky Way for the first time, and create a stellar dynamical model of the Milky Way's barred bulge. The model is constrained however by a previously constructed model of the Milky Way from \cite{BG02}, so this new model will be biased towards any inaccuracies from this previous model. The next generation of Milky Way model should be built directly from observational data of the Milky Way, and flexible enough for fitting heterogeneous data.

NMAGIC, developed by \cite{DeL07}, is the first algorithm to improve upon the initial M2M algorithm by adding the ability to include observational errors in the constraints. This is an important step forward as it allows real observational data to be used as constraints. NMAGIC was also the first M2M algorithm to use velocity constraints, in the form of line of sight spectra. NMAGIC has now been applied to several observed galaxies \citep[e.g.][]{DeL08,DeL09,DGMT11}. \cite{MG12} also made a recent improvement to the field of M2M modelling, by developing Moving Prior Regularization (MPR) which can replace the Global Weight entropy Regularization (GWR). \cite{MG12} show that MPR is beneficial to accuracy and smoothness in phase space distributions, and in some circumstances can converge to a unique solution, independent of the choice of the initial model. To examine M2M's performance against previous methods, \cite{LM12} have performed a direct comparison between M2M and the better known Schwarzchild method with regard to calculating the mass to light ratios of several elliptical and lenticular galaxies. 

These previous M2M algorithms use a distribution function or binned density distribution. However, the data that Gaia and the related surveys return will be in the form of individual stellar data. Therefore we have designed a particle-by-particle M2M algorithm that compares the observables at the location of each star (or the target particle) with the model observables at the same locations, and adjusts the weights in the same fashion as the original algorithm from \cite{ST96}. In this paper, we present proof of concept of the particle-by-particle M2M by recreating disc galaxies, generated with a Tree $N$-Body code, GCD+ \citep{KG03}. Our algorithm uses a self-consistent potential, which evolves over time along with the particle weights. We also show a model constructed from a partial target data set, demonstrating that the observables of the target galaxy do not have to cover the whole galaxy for M2M to work. This is the first step towards the real observational data from Galactic surveys, where the information will be provided for a limited region of the sky, with a more complicated selection function due to the dust extinction, crowding and stellar populations. The paper is organised as follows. Section \ref{M2M} describes the traditional M2M method and Section \ref{p2p} describes the methods behind our particle based adaptation. Section \ref{R} shows the performance of the particle-by-particle M2M for recreating the target disc system. In Section \ref{SFW} we discuss the accomplishments of this paper, and describe the next stages of our work.

\section{The M2M Algorithm}
\label{M2M}

In this section, we will give a brief description of the M2M algorithm as detailed in \cite{ST96},  \cite{DeL07} and \cite{LM10}, which forms the base for our work. The M2M algorithm works by calculating observable properties (observables hereafter) from the model and the target, and then adapting particle weights such that the properties of the model reproduce those of the target. The target can be in the form of a distribution function, an existing simulation, or real observational data. The model is always an $N$-body system.

The observables of the target system are described by
\begin{equation}
Y_{j}=\int{K_{j}(\textbf{z})f(\textbf{z})d^{6}\textbf{z}},
\label{tobs}
\end{equation}
where $j$ represents each individual observable, $\textbf{z}=(\textbf{r},\textbf{v})$ are the phase space coordinates, $f(\textbf{z})$ is the distribution function of the target galaxy and $K_{j}$ is a known kernel. Observables can come in many forms, including surface or volume densities, surface brightness and line of sight kinematics. The corresponding observable for the model takes the form
\begin{equation}
y_{j}=\sum_{i=1}^{N}w_{i}{K_{j}[\textbf{z}_{i}(t)]},
\label{mobs}
\end{equation}
where $w_{i}$ are the particle weights and $\textbf{z}_{i}$ are the phase space coordinates of the model's $i$-th particle.
We then calculate the difference in the observables of the target and the model, 
\begin{equation}
\Delta_{j} = \frac{y_{j}(t) - Y_{j}}{Y_{j}}.
\label{obsdiff}
\end{equation}
We then use this $\Delta_{j}$ to determine the so called force of change with the equation
\begin{equation}
\frac{d}{dt}w_{i}(t) = - \epsilon w_{i}(t)\sum_{j} \frac{K_{j}[\textbf{z}_{i}(t)]}{Z_{j}}\Delta_{j}(t),
\label{WC1}
\end{equation}
where $Z_{j}$ so far is an arbitrary constant, and the factor $K_{i}/Z_{j}$ can be thought of as the degree to which the $i$-th particle contributes to the $j$-th observable. $\epsilon$ is a parameter enabling us to control the rate of change. The linear dependence of equation (\ref{WC1}) upon $w_{i}$, coupled with the provision that a small enough $\epsilon$ is used, ensures that the weights do not become negative. \cite{ST96} show a proof of convergence for equation (\ref{WC1}) providing that the system starts close to the target.

If $N \textgreater J$, i.e. the number of the model particles, $N$, greatly exceeds the quantity of available constraints, $J$, the differential equation (\ref{WC1}) is ill-conditioned. \cite{ST96} suggest removing this ill conditioning by introducing entropy, by maximising the function
\begin{equation}
F = \mu S - \frac{1}{2}\chi^{2},
\label{Feq}
\end{equation}
where
\begin{equation}
\chi^{2} = \sum_{j} \Delta_{j}^{2},
\label{chi2}
\end{equation}
and $\mu$ is a parameter to control the regularization. The entropy is given by
\begin{equation}
S = - \sum_{i} w_{i} \ln\left(\frac{w_{i}}{\hat{w_{i}}}\right),
\label{S}
\end{equation}
where $\hat{w_{i}}$ are the priors, a predetermined set of weights, normally identical to each other such that $\hat{w}_{i}=M/N$, where $M$ is the total mass of the system and $N$ is the number of particles. The system can been normalised \citep{DeL07} such that
\begin{equation}
\sum_{i=1}^{N}w_{i} = 1.
\label{We1}
\end{equation}
This is useful if the total mass of the target system is one of the constraints. We do not impose this restriction as we wish to be able to create a system with a different total mass from the initial model.

Once the new entropy term is introduced to the force of change, equation (\ref{WC1}) is replaced by
\begin{equation}
\frac{d}{dt}w_{i}(t) = - \epsilon w_{i}(t)\left[\sum_{j} \frac{K_{j}[z_{i}(t)]}{Y_{j}}\Delta_{j}(t) - \mu\frac{\delta S}{\delta w_{i}}(t)\right],
\label{WC2}
\end{equation}
or
\begin{eqnarray}
\frac{d}{dt}w_{i}(t) = &-&\epsilon w_{i}(t)\Biggl[\sum_{j} \frac{K_{j}[z_{i}(t)]}{Y_{j}}\Delta_{j}(t) \nonumber \\
&+& \mu \left(\ln \left(\frac{w_{i}(t)}{\hat{w}_{i}}\right)+1\right)\Biggr],
\label{WC3}
\end{eqnarray}
for the most complete form. Note that $Z_{j}$ has been replaced by $Y_{j}$ due to the maximisation of equation (\ref{Feq}).

It is shown in \cite{ST96} and \cite{DeL07} that fluctuations in equation (\ref{obsdiff}) may be reduced by employing temporal smoothing, effectively boosting $N$ without drastically increasing computation time. This is achieved by replacing $\Delta_{j}(t)$ in equation (\ref{WC1}) with $\tilde{\Delta}_{j}(t)$, where
\begin{equation}
\tilde{\Delta}_{j}(t) = \alpha\int_{0}^{\infty}\Delta_{j}(t-\tau)\text{e}^{-\alpha\tau}d\tau,
\end{equation}
with $\alpha$ being small and positive. This $\tilde{\Delta}_{j}(t)$ can be calculated from the differential equation
\begin{equation}
\frac{d\tilde{\Delta}(t)}{dt} = \alpha(\Delta-\tilde{\Delta}).
\label{NewDel2}
\end{equation}
This temporal smoothing effectively increases the number of particles from $N$ to
\begin{equation}
N_{\text{eff}}=N\frac{t_{\frac{1}{2}}}{\Delta t},
\end{equation}
where $\Delta t$ is the length of the time step and $t_{\frac{1}{2}}=(\text{ln }2)/\alpha$ is the half life of the ghost particles. \cite{ST96} show that excessive temporal smoothing is undesirable, and should be limited to $\alpha\geq2\epsilon$.

The parameters $\epsilon$, $\mu$ and $\alpha$ must be determined via parameter search. We will discuss our choice of these parameters in Section \ref{PS}.

\section{Particle-by-particle M2M}
\label{p2p}
This section describes our original adaptation to the M2M algorithm. The majority of the methodology remains the same as described in Section \ref{M2M}, with the most substantial difference involving the Smoothed Particle Hydrodynamics (SPH) kernel \citep[e.g.][]{L77,GM77}, which will be described in Section \ref{p2pTheory}. \cite{ST96} used a kernel where they divide the coordinate space into bins. For example, for the density at the $j$-th bin with volume $V_j$, the kernel, $K_j(\textbf{r}_i$), is set to be $M_{tot}/V_j$ if $\textbf{r}_i$ is within the $j$-th bin, where $M_{tot}$ is the total mass of the system and equation (\ref{We1}) is satisfied. If $\textbf{r}_i$ is outside the $j$-th bin, $K_j(\textbf{r}_i)=0$. Because $K_j(\textbf{r}_1)$ and $K_j(\textbf{r}_2)$ are the same if $\textbf{r}_{1}$ and $\textbf{r}_{2}$ are in the same bin, this limits the resolution to the bin size. However as mentioned in Section \ref{intro-sec}, our ultimate target is the Milky Way, and the observables are not binned data, but the position and velocity of the individual stars which are distributed rather randomly. To maximise the available constraints, we evaluate the observables at the position of each star and compare them with the $N$-body model, i.e. in a particle-by-particle fashion. To this end we introduce a kernel often used in SPH, $W(r,h)$, which is a spherically symmetric spline function given by

\begin{equation}
\begin{array}{l}
W(r,h) = \frac{8}{\pi h^{3}} 
 \times \left\{ \begin{array}{cc}
 1-6(r/h)^{2}+6(r/h)^{3} & {\rm if}\ 0\leq r/h\leq 1/2,  \\
 2[1-(r/h)]^{3}      & {\rm if}\ 1/2\leq r/h\leq 1,  \\
 0               & {\rm otherwise}.
\end{array} \right.\\
\end{array} 
\label{Weq}
\end{equation}
as shown in \cite{ML85}, where $r = \mid\textbf{r}_i-\textbf{r}_j\mid$. Note that in our particle-by-particle M2M the kernel, $W(r,h)$, does not explicitly include the total mass, $M_{tot}$, because we wish to eventually apply it to the Milky Way, whose mass is unknown. Therefore the SPH kernel in equation (\ref{Weq}) is not equivalent to the M2M kernel, $K_j$, in Section 2.

Below, we describe our particle-by-particle M2M, considering that the target system is an $N$-body system whose particle position and velocity are known without any error. Of course in the real data of the Galaxy, there are complicated observational errors and selection functions, which often depend on stellar population and dust extinction. In this paper, we ignore these and consider an idealised system for a target. As described in Section \ref{intro-sec}, the aim of this paper is to demonstrate how our new M2M works and the potential of future application to the Galactic disc. We below assume that the target system consists of a single population, which we shall refer to as particles, and whose position and velocity are known without errors.

\subsection{Method}
\label{p2pTheory}
 We use the kernel of equation (\ref{Weq}) to calculate the density at the target particle locations, $\textbf{r}_j$, of both the target and the M2M model. Hereafter we replace the particle weights, $w_i$, with their masses $m_i$ due to our adoption of self-gravity in the particle-by-particle M2M. For example, the density of the target at $\textbf{r}_j$ is evaluated by,
\begin{equation}
\rho_{t,j}=\sum_{k=1}^{N}m_{t,k}W(r_{kj},h_{j}),
\label{trho}
\end{equation}
where $m_{t,k}$ is the mass of the target particle, $r_{kj} = \mid \textbf{r}_{k} - \textbf{r}_{j} \mid$, and $h_{j}$ is the smoothing length determined by 
\begin{equation}
h_{j} = \eta \left(\frac{m_{t,j}}{\rho_{t,j}}\right)^{1/3},
\label{smoothing}
\end{equation}
where $\eta$ is a parameter and we have set $\eta=3$. In SPH simulations, a value of $\eta$ between 2 and 3 are often used, and we employ the relatively higher value to maximise the smoothness. The solution of equation (\ref{smoothing}) is calculated iteratively until the relative change between two iterations is smaller than $10^{-3}$ \citep{PM07}. Similarly,
\begin{equation}
\rho_{j}=\sum_{i=1}^{N}m_{i}W(r_{ij},h_{j}),
\label{rho}
\end{equation}
from the model particles. The target density $\rho_{t,j}$ is calculated only once at the beginning of the M2M simulation, and the model density $\rho_{j}$ is recalculated at every timestep. 

For velocity constraints, we define the following form of the observables, using the same kernel. For example for radial velocity;
\begin{equation}
\delta v_{t,r,j}=\sum_{k=1}^{N}({v}_{t,r,k}-v_{t,r,j})m_{t,k}W(r_{kj},h_{j}),
\label{trv}
\end{equation}
where $v_{t,r,k}$ is the radial velocity of the $k$-th target particle and $v_{t,r,j}=(v_{t,x,j}x_{t,j}+v_{t,y,j}y_{t,j})/(x_{t,j}^2+y_{t,j}^2)^\frac{1}{2}$ is the radial velocity of the target system. Equation (\ref{trv}) represents the weighted mean of the relative velocities of the target particles within $h_j$ of the target particle $j$.
\begin{equation}
\delta v_{r,j}=\sum_{i=1}^{N}(v_{r,i}-{v}_{t,r,j})m_{i}W(r_{ij},h_{j})
\label{rv}
\end{equation}
is similarly calculated from the model particles. The same format is applied for the vertical and rotational velocities.

We then describe the difference in the observables i.e. equation (\ref{obsdiff}). For density;
\begin{equation}
\Delta_{\rho_j} = \frac{\rho_{j}(t) - \rho_{t,j}}{\rho_{t,j}}.
\label{od2}
\end{equation}
For velocity, we normalised them by the target density because of the density dependence introduced in equations (\ref{trv}) and (\ref{rv}), and therefore for the radial case;
\begin{equation}
\Delta_v=\frac{\delta v_{r,j}(t)-\delta v_{t,r,j}}{\sigma_{v_{r}} \rho_{t,j}}.
\label{Vdiff}
\end{equation}
Note that $\sigma$ is not an observational error, but just a normalisation constant which we have arbitrarily set to $\sigma_{v_r}=\sigma_{v_z}=\sigma_{v_rot}=10 \text{ km s}^{-1}$ in our demonstration in Section \ref{R}.

Because $\Delta_{\rho_j}$ and $\Delta_{v_j}$ are normalised differently, we modified their contribution to the force of change by introducing a new parameter $\zeta$ such that for our simulations, equation (\ref{WC3}) becomes, with smoothed $\tilde{\Delta}$ by equation (\ref{NewDel2});
\begin{eqnarray}
&\frac{d}{dt}m_{i}(t)& =  - \epsilon m_{i}(t)\Biggl[M\sum_{j} \frac{W(r_{ij},h_j)}{\rho_{t,j}}\tilde{\Delta}_{j,\rho}(t) \nonumber \\ 
&+& \zeta M\Biggl(\xi_r\sum_j\frac{W(r_{ij},h_j)}{\sigma_{v_{r}}\rho_{t,j}}(v_{r,i}-v_{t,r,j})\tilde{\Delta}_{v_{r,j}}(t) \nonumber \\
&+& \xi_z\sum_j\frac{W(r_{ij},h_j)}{\sigma_{v_{z}}\rho_{t,j}}(v_{z,i}-v_{t,z,j})\tilde{\Delta}_{v_{z,j}}(t) \nonumber \\
&+& \xi_{rot}\sum_j\frac{W(r_{ij},h_j)}{\sigma_{v_{rot}}\rho_{t,j}}(v_{rot,i}-v_{t,rot,j})\tilde{\Delta}_{v_{rot,j}}(t)\Biggr) \nonumber \\
&+& \mu \left(\ln \left(\frac{m_{i}(t)}{\hat{m}_{i}}\right)+1\right) \Biggr],
\label{WC4}
\end{eqnarray}
where $M$ is an arbitrary constant mass, which we set as $M=10^{12}$ M$_{\sun}$ for this paper. Note that in equation (\ref{WC4}) the corresponding M2M kernel is $K_j=MW(r,h_j)$, e.g. for density, which is inconsistent with the one used to obtain the observables in equation (\ref{rho}), where $K_j=M_{m,tot}W(r,h_j)$ and $M_{m,tot}$ is the total mass of the model particles. However we accept this inconsistency to apply the method to a system whose total mass is unknown and we allow $M_{m,tot}(t)=\sum_im_i(t)$ to freely evolve. Therefore we introduce the arbitrary constant $M$ in equation (\ref{WC4}), and as a result the parameters, such as $\epsilon$, $\mu$ and $\zeta$, must be calibrated for the specific system. Fortunately our ultimate target is only one system, the Milky Way. We hope that we can calibrate the parameters by modelling simulated data before applying the method to the real data. Hence note that the parameters presented in this paper are specific to the target system in this paper. In future works we will calibrate the parameters and refine the methods by applying more realistic simulation data.

 We use the additional individual parameters $\xi_{r}$, $\xi_{z}$, $\xi_{rot}$ for the different velocity observables, to allow us to fine tune their contributions to the force of change even further. Similar in spirit to \cite{DeL07}, we write $\epsilon$ as $\epsilon = \epsilon'\epsilon''$ where $\epsilon''$ is given by 
\begin{equation}
\epsilon'' = \frac{10}{\text{max}_{i}\left(M\sum_{j}\frac{W(r_{ij},h_j)}{\rho_{j,t}}\tilde{\Delta}_{\rho_j}(t)\right)},
\label{e''}
\end{equation}
for the density observable only.

In the previous works \citep[e.g.][]{ST96,Deh09,LM10,MG12}, the M2M method is applied to a system in a known fixed potential, i.e. using test particles. \cite{DeL07} demonstrate that M2M works with a partially self-consistent potential, in that the potential is calculated every 25 time steps, setting the particle mass $m_i=w_iM_{tot}$. However this repeated sudden change of the potential could come with some problems that will be discussed later. 

We intend to apply our algorithm to the Milky Way, whose mass distribution is poorly known \citep[e.g.][]{Mc11}, and one of the aims of applying the dynamical model is to reconstruct the mass distribution. Therefore we use a self-consistent disc potential, setting the particle weight, $w_{i}$, to the mass, $m_{i}$, allowing the disc potential to change along with the model observables and allowing us to recover simultaneously the disc potential along with the mass and velocity profiles. In this paper we focus on the disc. We ignore the bulge or halo stars, and assume that the dark matter potential is known for this initial demonstration. Note that the previous studies are mainly focused on elliptical galaxies, i.e. systems dominated by velocity dispersion, but not strongly rotation supported. Recreating a disc galaxy with a self-consistent potential has been attempted once before by \cite{DegPhD}, who highlights some difficulties with an M2M method that employs self-gravity. He uses a grid to calculate the observables, which makes his method different from ours. 

One of the problems arising from using a self-consistent potential as mentioned by 
\cite{DegPhD} is that the temporal smoothing, which worked well in fixed potential M2M methods, is problematic when used with self-gravity. The temporal smoothing reduces shot noise by averaging the $\Delta_j$ back along their orbits, which is fine with test particles in a fixed potential because the orbits are fixed. However in a self-consistent potential, the potential and therefore particle orbits change with time, and thus the temporal smoothing breaks the self-consistency. Therefore we should  be aware that self-gravity M2M models are very sensitive to instabilities, and we see substantial disruption when the smoothing is first turned on. A way to mitigate this damage due to the temporal smoothing is described in Section \ref{p2pSetup}. In light of this we investigated the possibility of running models without temporal smoothing. However all models had to be substantially under-regularized to recover the velocity profiles shown in Section \ref{R}, which leads to the continuous fluctuation of the weights, similar to the problems of the under-regularization discussed in Section \ref{Smu}.

We use a standard Euler method for the integration of the weight change equation and a leapfrog time integrator for advancing the particles. We also use individual time steps for the particles, and only update the masses of particles whose position and velocity are updated within the individual timestep. The timestep for each particle is determined by
\begin{equation}
dt_i=C_{\text{DYN}}\left(\frac{0.5h_i}{\vert d\textbf{v}_i/dt\vert}\right)^{\frac{1}{2}},
\end{equation}
with $C_{\text{DYN}}=0.2$.

\subsection{Target System Setup}
\label{p2pSetup}
Our simulated target galaxy consists of a pure stellar disc with no bulge and a static dark matter halo, set up using the method described in \cite{GKC12}. The dark matter halo density profile is taken from \cite{NFW97};
\begin{equation}
\rho_{dm}=\frac{3H_0^2}{8\pi G}\frac{\delta_c}{cx(1+cx)^2},
\end{equation}
where $\delta_c$ is the characteristic density described by \cite{NFW97}. The concentration parameter $c=r_{200}/r_s$ and $x=r/r_{200}$, where $r_{200}$ is the radius inside which the mean density of the dark matter sphere is equal to $200\rho_{crit}$ and given by;
\begin{equation}
r_{200}=1.63\times10^{-2}\left(\frac{M_{200}}{h^{-1}M_{\sun}}\right)^{\frac{1}{3}}h^{-1}\text{kpc}.
\end{equation}
We use $M_{200}=1.75\times10^{12}$ M$_{\sun}$, $c=20$ and $H_0=71\text{ km s}^{-1}\text{Mpc}^{-1}$.

\begin{figure}
\centering
\resizebox{\hsize}{!}{\includegraphics{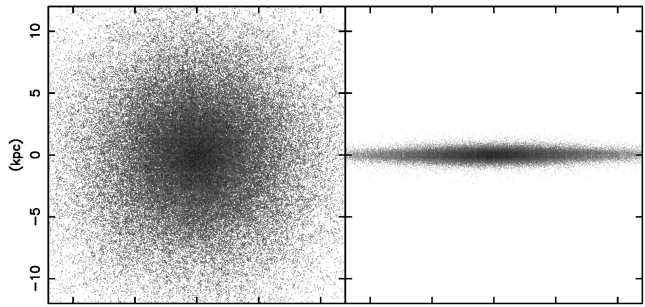}}
\caption{The end result ($t=2$ Gyr) of an $N$-body disc galaxy simulation. It had a scale length of 3 kpc initially. This will be used as the target system as shown in the Section \ref{R}. The left and right panels show the face-on and end-on views respectively.} 
\label{HHH}
\end{figure}

The stellar disc is assumed to follow an exponential surface density profile:
\begin{equation}
\rho_d=\frac{M_d}{4\pi z_dR_d^2}\text{sech}^2\left(\frac{z}{z_d}\right)\text{e}^{-R/R_d},
\end{equation}
where $z_d$ is the scale height of the disc and $R_d$ is the scale length. Our target disc has $z_d=0.35\text{ kpc}$ and $R_d=3.0\text{ kpc}$. The disc has a mass of $M_d=3.0\times10^{10}$ M$_{\sun}$ and consists of $10^5$ particles, with each particle having a mass of $3.0\times10^{5}$ M$_{\sun}$. We use the kernel softening suggested by \cite{PM07}. Although \cite{PM07} suggested adaptive softening length, we use a fixed softening for these simulations for simplicity. Our definition of the softening length $\varepsilon=1.05\text{ kpc}$ is about three times larger than the equivalent Plummer softening length. We also use this for the M2M modelling runs. The velocity dispersion for each three dimensional position of the disc is computed following \cite{SMH05} to construct an almost equilibrium condition. We use a high value of the free parameter $f_R=\sigma_R/\sigma_z=3$, which controls the ratio between the radial and vertical velocity dispersions, to deliberately suppress structure formation and create a smooth, almost axisymmetric disc for this initial test. Our target system is a relatively smooth disc galaxy evolved over 2 Gyr, as shown in Fig. \ref{HHH}, and it is used for all models in Section \ref{R}. 

\begin{table*}
\centering
\caption{M2M model parameters}
\label{pars}
\renewcommand{\footnoterule}{}
\begin{tabular}{@{}cccccccccccc@{}}
\hline
Model & $R_{d,ini}$ (kpc) & $\epsilon'$ & $\mu$ & $\alpha$ & $\zeta$ & $\chi^2_\rho$ & $\chi^2_{v_r}$ &  $\chi^2_{v_z}$ &  $\chi^2_{v_{rot}}$ & Notes\\
\hline
A & 2.0 & 0.1 & $5\times10^{5}$ & $0.2$ & $0.05$ & 0.0846 & 7.291 & 0.918 & 6.502 & Fiducial\\ \hline
B & 2.0 & 0.1 & $10^{4}$ & $0.2$ & $0$ & 0.0831 & 9.599 & 1.074 & 10.873 & No Velocity\\ \hline
C & 2.0 & 0.1 & $10^{4}$ & $0.2$ & $0.05$ & 0.0912 & 8.275 & 1.069 & 7.464\\ \hline
D & 2.0 & 0.1 & $10^{5}$ & $0.2$ & $0.05$ & 0.0875 & 7.914 & 1.005 & 7.087\\ \hline
E & 2.0 & 0.1 & $10^{6}$ & $0.2$ & $0.05$ & 0.0894 & 7.099 & 0.893 & 6.440\\ \hline
F & 2.0 & 0.1 & $10^{7}$ & $0.2$ & $0.05$ & 0.223 & 9.395 & 1.130 & 9.960\\ \hline
G & 2.0 & 0.1 & $10^{8}$ & $0.2$ & $0.05$ & 0.407 & 17.291 & 2.107 & 17.701\\ \hline
H & 5.0 & 0.1 & $5\times10^5$ & $0.2$ & $0.05$ & 0.100 & 10.839 & 1.414 & 10.394\\ \hline
I & 6.0 & 0.1 & $5\times10^5$ & $0.2$ & $0.05$ & 0.111 & 12.381 & 1.604 & 12.094\\ \hline
J & 1.5 & 0.1 & $5\times10^5$ & $0.2$ & $0.05$ & 0.101 & 7.849 & 0.972 & 6.896\\ \hline
K & 2.0 & 0.1 & $5\times10^{5}$ & $0.2$ & $0.05$ & 0.0924 & 7.309 & 0.911 & 6.509 & Partial Data\\ \hline
\end{tabular}
\end{table*}

\begin{figure}
\centering
\resizebox{\hsize}{!}{\includegraphics{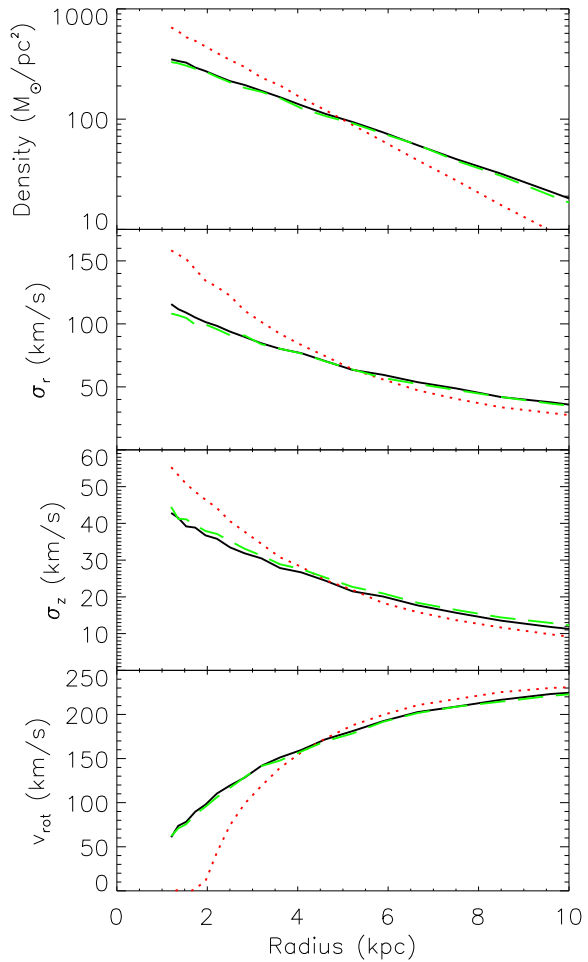}}
\caption{Initial (red dotted), final (green dashed) and target (black solid), density profile (upper), radial velocity dispersion (upper middle), vertical velocity dispersion (lower middle) and rotational velocity (lower) for Model A. The initial model has a scale length of 2 kpc, the target model has a scale length of 3 kpc.} 
\label{GrA}
\end{figure}

We set up the initial conditions of the model disc with the same parameters and method, but use a different scale length from that of the target galaxy.

\subsection{Procedure}
\label{Procedure}
The sudden change in potential caused by the changing particle weights induces instabilities and potentially unwanted structure formation. This effect can be reduced by dividing the modelling process into a series of stages, each with a slightly different level of M2M algorithm. This reduces the magnitude of the change in potential at any one time. We also set a limit on the maximum change in mass any particle can experience in one time step. We set this limit to ten percent of that particles mass. 

Initially the model is allowed to relax in a pure self-gravity environment with no M2M constraints for 0.471 Gyr (our $N$-body code time unit). This relaxation period is important, as applying the M2M algorithm before the model has settled generates the aforementioned instabilities. Although our M2M algorithm was still capable of recovering the desired profile, the time scale needed was drastically increased because the model had to smooth out again before convergence took place if we turned on the M2M without the relaxation period.

After this period of relaxation, the M2M algorithm is activated and runs without temporal smoothing for a further 1.413 Gyr, which allows the density and velocity profiles to converge quickly. During this time, the contribution of the velocity constraint is increased linearly from 0 up until our desired $\zeta$. This allows the density profile to converge first. We found this slow increase in the velocity constraints to be important, because if the velocity constraints were introduced simultaneously at full strength, we find the large weight changes induce the sudden potential change mentioned earlier, which is strong enough to disrupt the disc.

Then, after 1.884 Gyr, the temporal smoothing is turned on. When the M2M modelling was run with temporal smoothing from the beginning, the mass profile experienced large oscillations. The modelling then continues in this state for as long as is desired. Our M2M models are run for a period of 10 Gyr.

\subsection{Parameter Calibration}
\label{PS}
As discussed in \cite{ST96}, \cite{DeL07}, \cite{LM10} and \cite{MG12}, the choice of parameters are crucial for the success of M2M modelling. In this section, we will discuss our choice of the parameters; $\epsilon$, $\alpha$, $\zeta$ and $\mu$, and how we calibrate these values. Note that these parameters are calibrated for this specific target system. It is likely that we need different calibration for different targets. However what we learned from the parameter search should be useful for future applications and developments of the improved version.

$\epsilon$ provides the balance between the speed of convergence, and the smoothness of the process. In this case, we find that when $\epsilon' > 0.1$, the weights change too rapidly,  which induces the sudden potential changes and therefore more instabilities. This leads to a general decrease in the final level of accuracy of both density and velocity profiles. If $\epsilon'\leq0.1$ convergence can be achieved and the particle weights experience a much smoother evolution. However if $\epsilon'$ is too small, the oscillations generated by the temporal smoothing take too long to damp down, which drastically increases the length of the simulation. In the end, we have chosen $\epsilon'=0.1$ as a balance between accuracy and simulation time. With more computing power available to us we would consider running a lower value of $\epsilon$. However if $\epsilon'\ll0.1$, it is possible the model will not show any signs of convergence as the weight change is too slow. 

The choice of $\alpha$, which controls the strength of the temporal smoothing, should depend upon the choice of $\epsilon$ ($\alpha\geq2\epsilon$). Note that $\epsilon=\epsilon'\epsilon''$ and $\epsilon''$ is defined by equation (\ref{e''}). We find that our modelling is not sensitive to $\alpha$ and we set $\alpha=0.2$ in this paper.

\begin{figure}
\centering
\resizebox{\hsize}{!}{\includegraphics{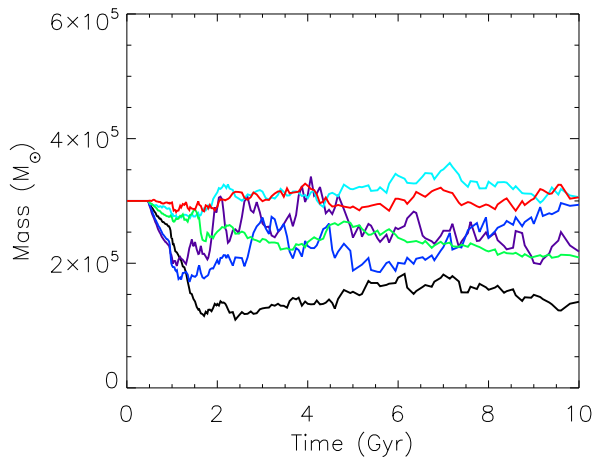}}
\caption{The weight evolution for a selection of particles from Model A.}
\label{IdA}
\end{figure}

$\zeta$ (and individual $\xi$) controls the level of the velocity constraints. It is important to strike a balance between the density and velocity constraints, because if the level of constraints are unbalanced one will dominate in the change of weight and the other observables will not converge. We can choose a suitable $\zeta$ (and/or $\xi_r,\xi_z$ and $\xi_{rot}$) by comparing the magnitudes of the individual terms of the right-hand-side of equation (\ref{WC4}). We set $\zeta$ such that the contribution of the velocity constraint to the force of change equation is the same magnitude as, or slightly less than, the density constraints. The individual velocity components may then be fine tuned with $\xi_j$. For our simulations, we find that the following parameter set works well: $\zeta=0.05$, $\xi_r=1$, $\xi_z=10$ and $\xi_{\textbf{rot}}=1$. 

$\mu$ controls the strength of the regularization. We discuss the importance of $\mu$ in greater detail in Section \ref{Smu}. In our fiducial model shown in Section \ref{R} we adopt $\mu=5\times10^5$.

\section{Particle-by-Particle M2M Results}
\label{R}

In this section we present the results from our modelling of our target disc galaxy. We will first show the results for our fiducial model, and then compare it with a model using only density constraints. We ran multiple M2M models with different parameters, which can be seen in Table \ref{pars}, where $R_{d,ini}$ is the initial scale length of the model disc. We only use the observables within the radius of $10\text{ kpc}$.

\subsection{Fiducial Model}
In this section we present Model A, our fiducial model constructed with the parameters described in Section \ref{PS}, and shown in Table 1. We start from an $N$-body disc with a scale length of 2 kpc, recreating the target disc ($R_{d}=3\text{ kpc}$) with our particle-by-particle M2M, evolving the model for 10 Gyr. Fig. \ref{GrA} shows the radial profiles of the surface density, radial and tangential velocity dispersion, and the mean rotational velocity. The final profiles of Model A reproduce the profiles of the target system remarkably well. Note that these radial profiles are not direct constraints of the particle-by-particle M2M. Especially it is rather surprising that the velocity dispersion profiles are recovered. We think that this is because the particle-by-particle M2M forces the model particles to follow the velocity distribution of the target particles. We also have no constraints on the total mass of the disc. Note also that the assumed velocity constant is $10 \text{ km s}^{-1}$ in equation (\ref{Vdiff}) yet the velocity profiles are reproduced at a level much less than $10 \text{ km s}^{-1}$. This is not surprising however, because we have different normalisations for density and velocity, and adjust $\zeta$ and $\xi$ to balance their contributions in equation (\ref{WC4}) making the choice of $\sigma$ arbitrary. Therefore, $\sigma_{v_r,t}$ is not indicating an error, but is merely a constant value for normalisation. In this paper, we do not include any error. We plan to add more realistic errors in future works. Fig. \ref{IdA} shows the weight evolution for a selection of particles from Model A. Weight convergence is adequate, however it is not as smooth as desired. We find that the particle weight evolution is less smooth for the case where velocity observables have been added.
\begin{figure}
\centering
\resizebox{\hsize}{!}{\includegraphics{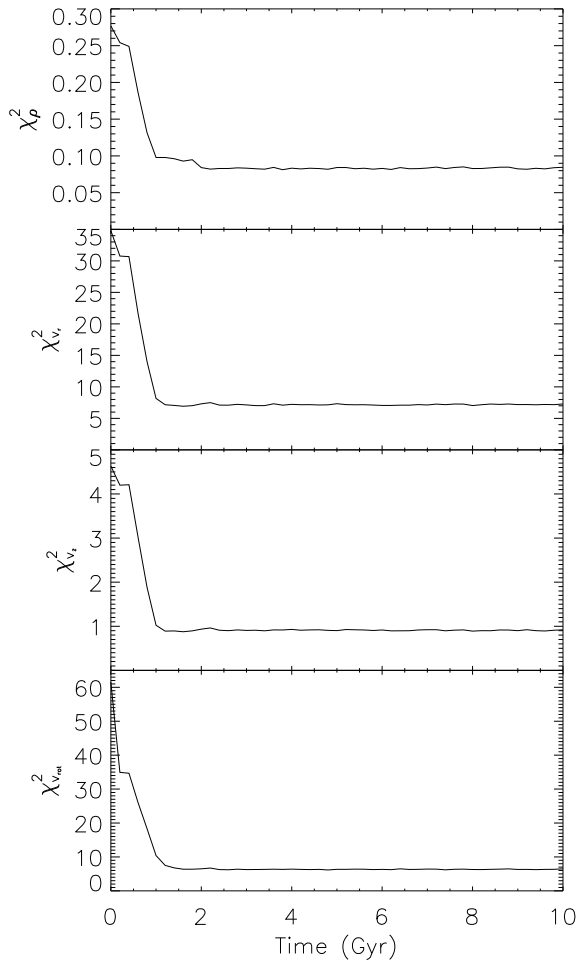}}
\caption{Time evolution of $\chi^2$ for density (upper), radial velocity (upper middle), vertical velocity (lower middle) and rotational velocity (lower) for Model A.}
\label{XiA}
\end{figure}
\begin{figure}
\centering
\resizebox{\hsize}{!}{\includegraphics{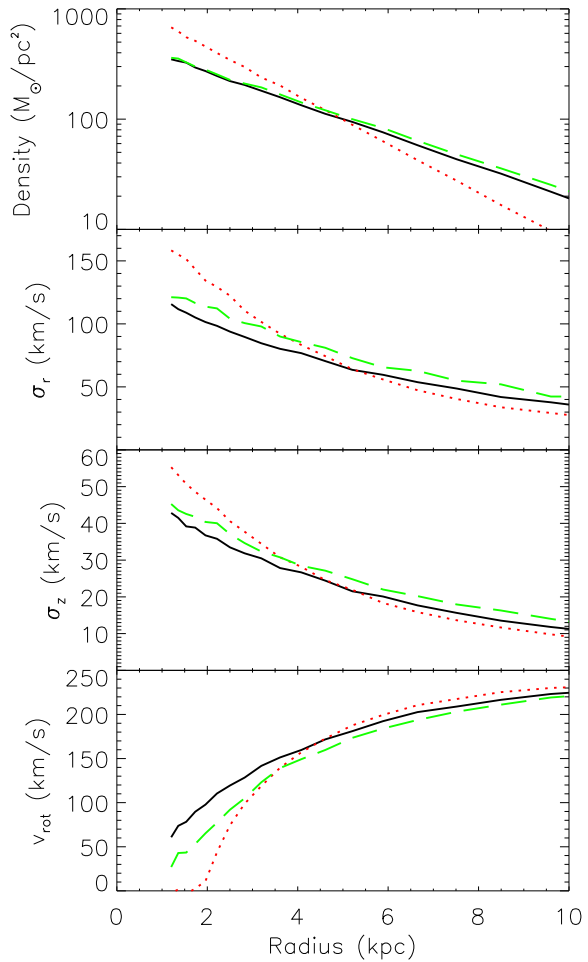}}
\caption{Same as Fig. \ref{GrA}, but for Model B which uses only the density observable as a constraint.} 
\label{GrB}
\end{figure}
Fig. \ref{XiA} shows the $\chi^2$ evolution for each of the observables. For all observables we use 
\begin{equation}
\chi^2_X=\frac{\sum\Delta_X^2}{N_r},
\end{equation}
where $\Delta_X$ is equivalent to equations (\ref{od2}), i.e. $X=\rho$, and (\ref{Vdiff}), i.e. $X=v$. This is a slightly unusual definition of $\chi^2$ for the velocity observables. Note that we include only particles within 10 kpc and $N_{r}$ is the number of target particles satisfying this criteria. In Model A, $\chi^2$ values rapidly decrease until 2 Gyr, from which point there is almost no improvement. The final values of $\chi^2$ are also shown in Table 1. 

In comparison we show Model B, with the same initial conditions and target with the velocity constraints turned off. We find that  $\mu=5\times10^5$ cause over-regularization for this case, and has to be reduced in compensation to $\mu=10^4$. Fig. \ref{GrB} shows the density and velocity profiles for Model B. The final model-density profile resembles the target. Due to the lack of velocity constraints, while the velocity profiles do improve, they do not resemble the target. A comparison between Fig. \ref{GrA} and Fig. \ref{GrB} demonstrates how the velocity constraints improve our reproduction of the dynamical properties of the target. 

\begin{figure}
\centering
\resizebox{\hsize}{!}{\includegraphics{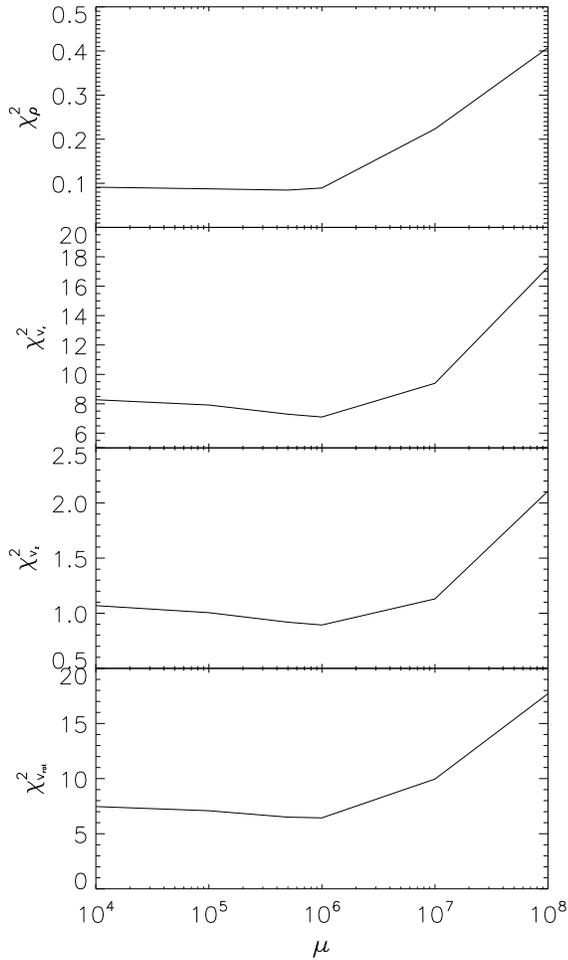}}
\caption{Accuracy of our final M2M model dependent on $\mu$ as determined by $\chi^2$ for density (upper), radial velocity (upper middle), vertical velocity (lower middle) and rotational velocity (lower).}
\label{Mu}
\end{figure}
\begin{figure}
\centering
\resizebox{\hsize}{!}{\includegraphics{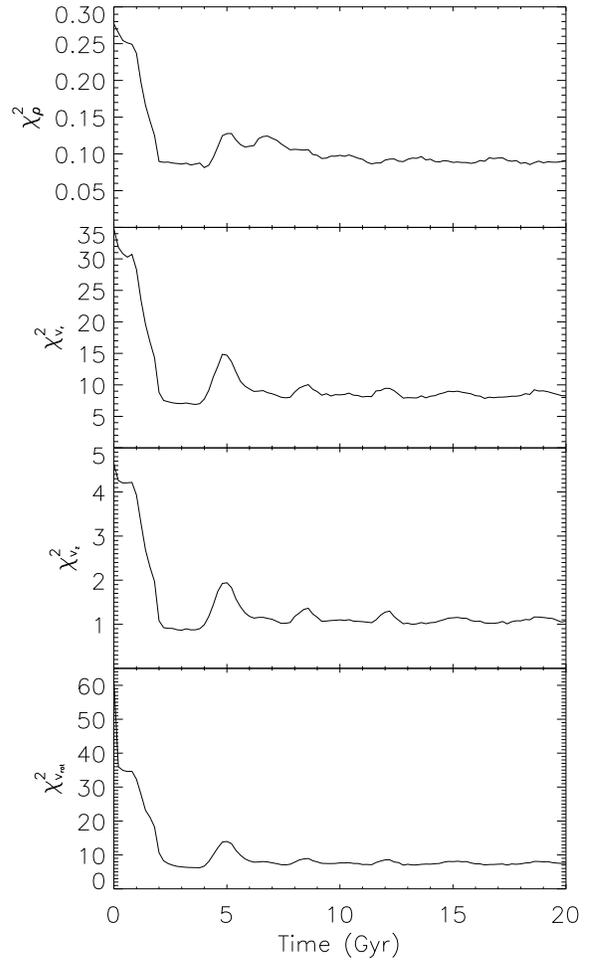}}
\caption{Same as Fig. \ref{XiA}, but for Model C, with $\mu=10^4$.}
\label{XiC}
\end{figure}

\subsection{Effect of Regularization}
\label{Smu}
Similar to the previous studies \citep[e.g.][]{ST96,DeL07,LM10}, we also find that careful choice of the value of $\mu$ is key to obtain convergence to a good model, and reproduce the given observables. Therefore we discuss in this section how $\mu$ affects the modelling. We performed multiple models with the same initial conditions and parameters as Model A, except the value of $\mu$ (see Models C-G in Table \ref{pars}). Fig. \ref{Mu} shows the $\chi^2$ for the density and velocity at the final time ($t=10$ Gyr). The figure demonstrates a slow improvement for the three velocity observables with an increasing $\mu$ up until a value of approximately $\mu=10^6$, above which goodness of fit drops off again. The density observable appreciates a slightly lower value of $\mu$.

\begin{figure}
\centering
\resizebox{\hsize}{!}{\includegraphics{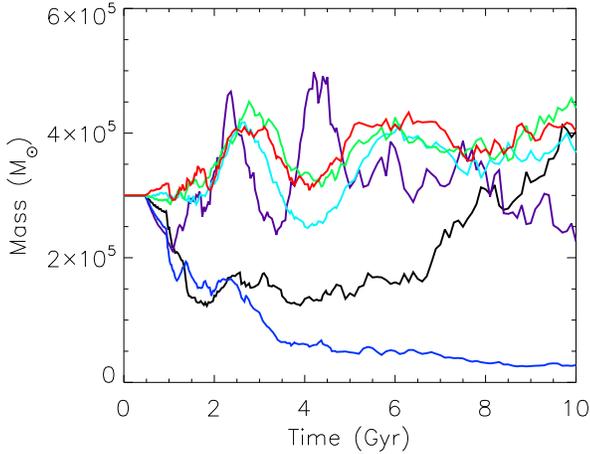}}
\caption{Same as Fig, \ref{IdA}, but for Model C, with $\mu=10^4$.}
\label{IdC}
\end{figure}

Although there is not a vast difference between the final values of $\chi^2$ for $\mu=10^4,10^5,10^6$, Model C with $\mu=10^4$ is found to be an inappropriate model because of its poor convergence. Fig \ref{XiC} shows the time evolution of $\chi^2$ in Model C. Fig. \ref{XiC} shows oscillatory behaviour. Fig. \ref{IdC} shows the time evolution of the weight for the particles selected in Fig. \ref{IdA}. Comparison between Figs. \ref{IdA} and \ref{IdC} demonstrates that $\mu=10^4$ is too low to suppress the large amplitude of the fluctuations in the particle weights. The weights of the particles keep changing and do not converge. Therefore we judge that $\mu=10^4$ is unacceptable for recreating the target system.

\begin{figure}
\centering
\resizebox{\hsize}{!}{\includegraphics{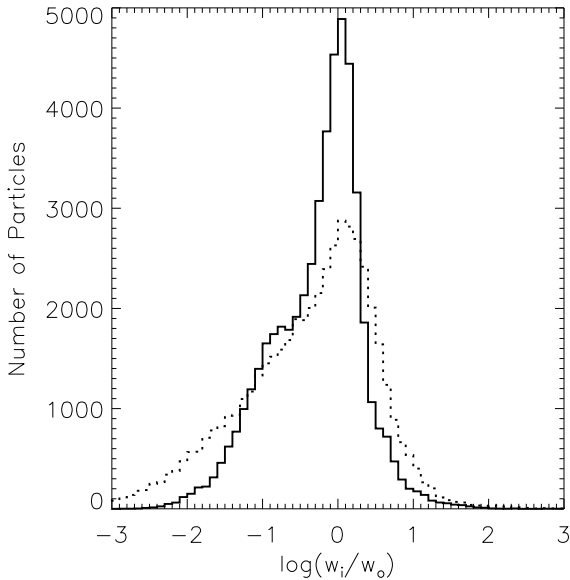}}
\caption{Distribution of particle weights for Model A (solid) and Model C (dotted) at the final time, $t=10$ Gyr. $w_0$ indicates the initial particle weights.}
\label{Histo}
\end{figure}

Fig. \ref{Histo} displays the distribution of particle weights at the final time for Models A and C. The histogram shows a wider tail, and lower peak for the under-regularized Model C compared to our fiducial Model A. This is expected because a higher $\mu$ restricts particles from moving far from the initial mass used as a prior. As a result, Model A shows a narrower distribution and thus a higher peak close to the initial value of $w_i$.
Fig. \ref{Histo} also demonstrates that $\mu=10^4$ is less favourable.

\begin{figure}
\centering
\resizebox{\hsize}{!}{\includegraphics{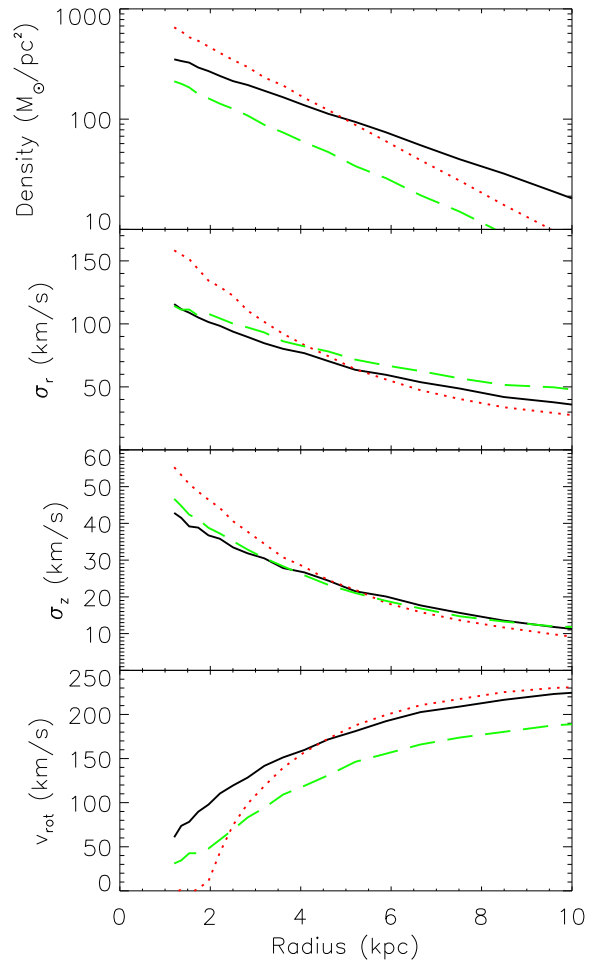}}
\caption{Same as Fig. \ref{GrA} but for Model G, with $\mu=10^8$.} 
\label{GrG}
\end{figure}

If we examine substantial over-regularization, i.e. a higher value of $\mu$, it is easy to see the damaging effect on the density and velocity profiles. Fig. \ref{GrG} shows the profiles from Model G, with $\mu=10^8$, which shows the significant discrepancy in the density and rotational velocity profiles between the final profiles and the target profiles. The discrepancy in the other two velocity observables is not as substantial. However it is clearly worse when compared with Fig. \ref{GrA}. 

In summary, we found that we required regularization of around $\mu=10^5-10^6$ as a compromise between the goodness of fit, and the smoothness of the $\chi^2$ and particle weight evolution. Both $\mu=10^7$ and $\mu=10^8$ show over-regularization and the density profiles associated with those values converge to an incorrect profile. $\mu=10^4$ shows large oscillations in both $\chi^2$ and particle weights, and convergence is not reached. Anything in the range of $\mu=10^5-10^6$ appears appropriate and hence our fiducial model adopts $\mu=5\times10^5$. As can be seen from Table \ref{pars}, we find under-regularization is preferable to over-regularization. This is also the case in previous literature \citep[e.g.][]{DeL08,MG12} implying this is a generic feature of M2M and not intrinsic to any specific algorithm.

\subsection{Different Initial Conditions}
\begin{figure}
\centering
\resizebox{\hsize}{!}{\includegraphics{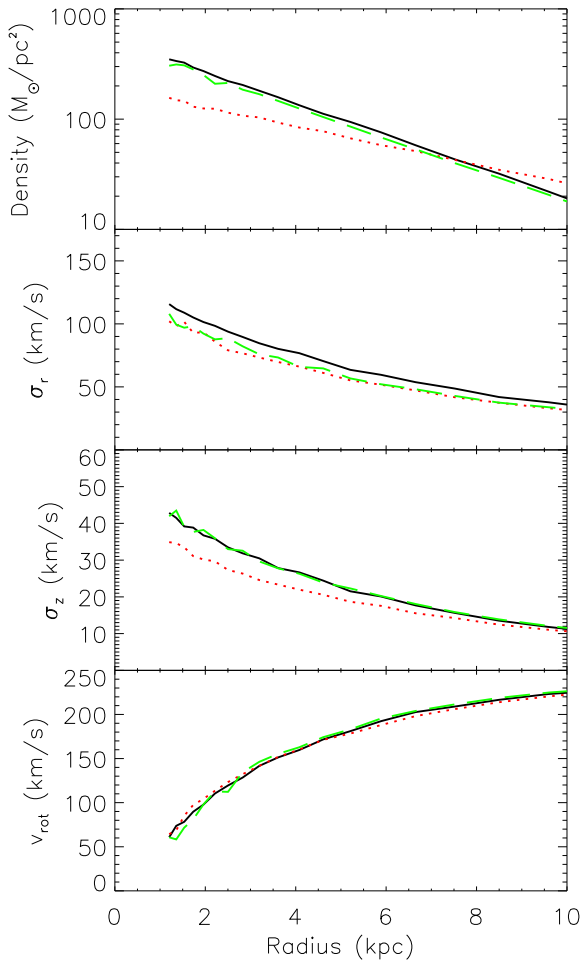}}
\caption{Same as Fig. \ref{GrA}, but for Model H, where $R_{d,ini}=5.0$ kpc.}
\label{GrH}
\end{figure}

\begin{figure}
\centering
\resizebox{\hsize}{!}{\includegraphics{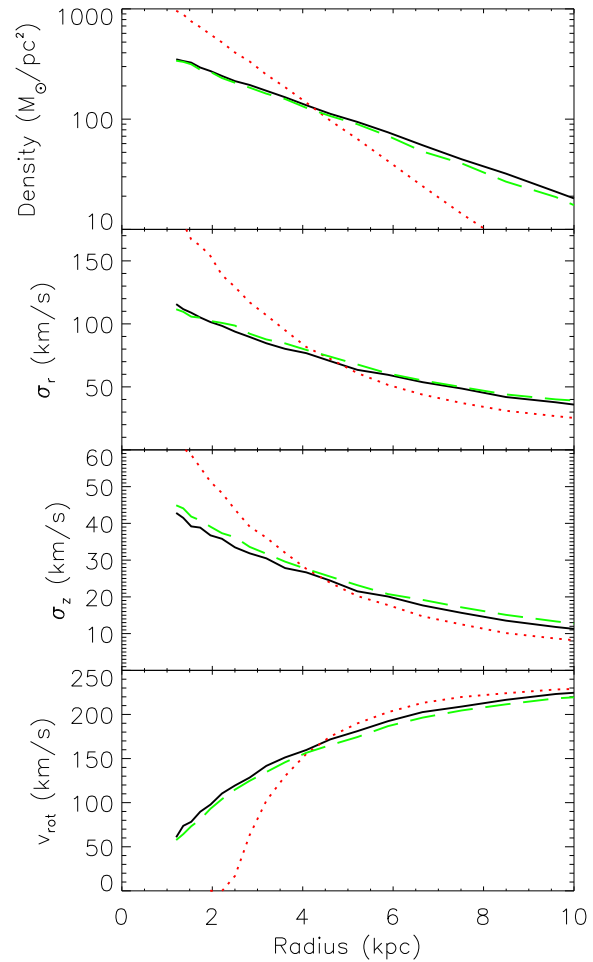}}
\caption{Same as Fig. \ref{GrA}, but for Model J, where $R_{d,ini}=1.5$ kpc.}
\label{GrJ}
\end{figure}

We also tested the algorithm on the same target, using initial discs with a different scale length, but with the same parameters as Model A. We have already discussed the benefits of tailoring $\mu$ to the model, so we were not expecting that these models (Models H-I in Table \ref{pars}) would recreate their target systems to the same level as Model A. However for demonstration purposes, we show how the parameter set in Model A works if the initial conditions are different. When we started from a higher initial scale length (Model H with $R_{d,ini}=5\text{ kpc}$ and Model I with $R_{d,ini}=6\text{ kpc}$) we attained a reasonable reproduction of the target, However, the final $\chi^2$ is systematically higher than Model A (see Table 1). Fig. \ref{GrH} shows the profiles from Model H, which slightly disagree with the targets. This seems to be due to over-regularization, and we would need to adjust $\mu$ in order to obtain a better model. On the other hand, when we started from a lower initial scale length (Model J with $R_{d,ini}=1.5$ kpc, the profiles are shown in Fig. \ref{GrJ}), $\chi^2$ was only fractionally worse than the fiducial case Model A (see Table \ref{pars}). This demonstrates that it is better to set the initial disc with a smaller scale length. In the application to the real observational data of the Milky Way, we do not know the right shape of ``the target model". However, we hope that the further studies with these target galaxies would help us to understand more about how the M2M modelling behaves in different cases, and how we should calibrate the parameters.
\begin{figure}
\centering
\resizebox{\hsize}{!}{\includegraphics{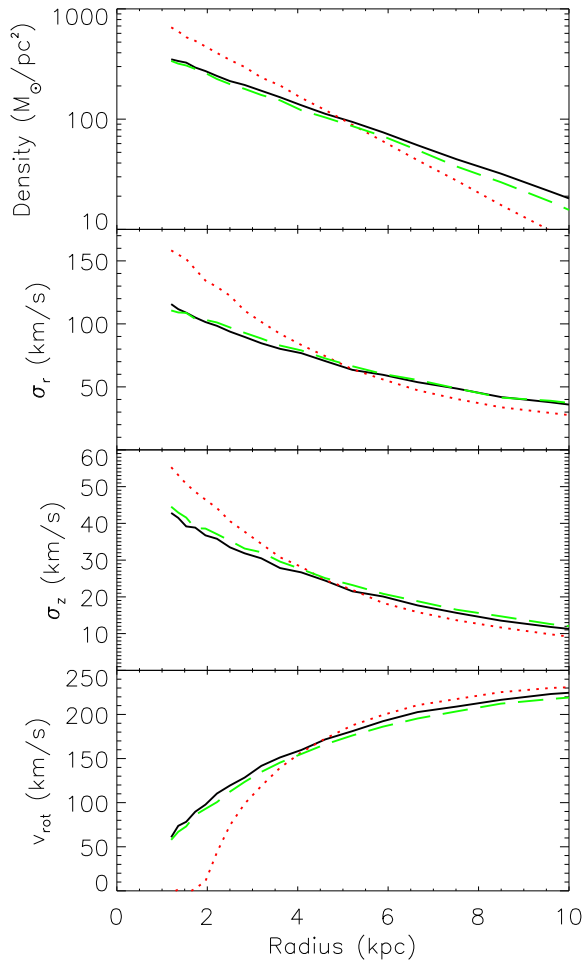}}
\caption{Same as Fig. \ref{GrA}, but for Model K, where the observables are calculated only in a sphere of 10 kpc around a point in the plane 8 kpc from the galactic centre.}
\label{GrK}
\end{figure}

\subsection{The Partial Data Case}
\label{OSph}
Because our goal is to eventually use our method with Gaia data, and Gaia will only survey a section of the Galactic disc, it is important to test our algorithm on an incomplete data set (Model K in Table \ref{pars}). In this paper, a simple selection function is applied for the purpose of demonstration. Remember that our models in the previous sections have used only the data within the radius of 10 kpc from the centre. In this section we additionally restricted the observables within a 10 kpc sphere around a point in the plane, 8 kpc away from the Galactic centre.

Fig. \ref{GrK} shows the final profiles for Model K, which reproduces the target profiles reasonably well. Compared with Fig. \ref{GrA}, Fig. \ref{GrK} shows only a minor discrepancy to the target profiles, mainly in the outer region. Worse performance in the outer region is unsurprising, as the larger the radii, the smaller percentage of the particles orbits are spent within the sampled area. Table \ref{pars} shows the final values of $\chi^2$, which displays a better value of $\chi^2$ than over-regularized models, and similar levels of the goodness of fit to under-regularized ones, without the excessive weight oscillations. Model K demonstrates that it is possible to apply our particle-by-particle M2M to a disc galaxy with only a limited selection of data.

\section{Summary and Further Work}
\label{SFW}
We have developed the initial version of our new particle-by-particle M2M, where the observables are compared at the position of the target particles, and the gravitational potential is automatically adjusted by the weight change of the particles. This paper demonstrates that the particle-by-particle M2M can recreate a target disc system in a known dark matter potential. The radial profiles of the surface density, velocity dispersion in the radial and perpendicular directions, and the rotational velocity of the target disc are all well reproduced from the initial disc model whose scale length is different from that of the target disc. We find that the regularization parameter, $\mu$, is key to obtaining a reasonable convergence to a satisfactory model. We also demonstrate that our M2M can be applied to an incomplete data set and recreate the target disc reasonably well when the observables are restricted to within a sphere of radius 10 kpc around a point in the disc plane 8 kpc from the centre.

Admittedly, these applications are simplified cases. Our ultimate goal is to develop the M2M to be applicable to the observational data that Gaia and other related Galactic surveys will provide. As discussed above, Gaia will produce an unprecedentedly large amount of data for the order of a billion stars, with many dimensions of information. The accuracy of each dimension of information could be quite inhomogeneous, depending on distance, stellar population, and location in the sky due to dust extinction, crowding etc. meaning that the observational selection function is quite complex. There are many challenges before us to develop the M2M for Gaia type data.

We believe that as shown in this paper, it is a good practice for Galaxy modelling to attempt to reconstruct galaxy models created by $N$-body simulations, where the full dimensions of the properties are known. Although as an initial attempt, we have taken a disc without any non-axisymmetric structure, we are trying to apply the method to $N$-body discs with spiral arms and a bar. In the future we will add more realistic errors and selection functions, to account for dust extinction and crowding. We must then take into account the expected Gaia performances, the stellar population \citep[e.g.][]{SBJB11,PCK12} and the three dimensional dust extinction models \citep[e.g.][]{DS01,MRRSP06}. We realise that the observables used in this paper are not ideal for such complicated data. We are also investigating other forms of observable and their associated contribution to the force of change, such as the maximum likelihood as applied in \cite{DeL08}. In this paper, we assume that the dark matter halo potential is known and spherical for simplicity. However, of course we do not know the shape of the dark matter of the Milky Way, and in reality we have to simultaneously explore the different dark matter potential. Another important question is the uniqueness of our M2M solution. Even if the M2M model explains all the observables similarly well. The question remains, what are the ``real" dynamical models for the Milky Way? We hope that many exercises with these ``fake" targets created by $N$-body simulations will be useful to identify the uniqueness of the obtained dynamical model and possible systematic biases.

Encouraged by the success of this paper, we are further improving our particle-by-particle M2M to apply them to the upcoming Galactic observational data, and ultimately construct a dynamical model of the Milky Way.

\section*{Acknowledgements}
We thank an anonymous referee for their careful review that improved the manuscript. We gratefully acknowledge the support of the UK's Science \& Technology Facilities Council (STFC Grant ST/H00260X/1). The calculations for this paper were performed on Cray XT4 at Center for Computational Astrophysics, CfCA, of the National Astronomical Observatory of Japan and the DiRAC Facility jointly funded by the STFC and the Large Facilities Capital Fund of BIS. The authors also thank the support of the STFC-funded Miracle Consortium (part of the DiRAC facility) in providing access to the UCL Legion High Performance Computing Facility. We additionally acknowledge the support of UCLs Research Computing team with the use of the Legion facility. JH would also like to thank Richard Long for a constructive discussion about M2M, and Kirthika Mohan for proof reading the manuscript.

\bibliographystyle{mn2e}
\bibliography{ref2}

\label{lastpage}
\end{document}